\documentclass[12pt]{article}
\usepackage{amsfonts}
\usepackage{amsmath}
\usepackage{amssymb}
\usepackage{mathptmx}
\usepackage[dvips]{color}
\usepackage{epsfig}
\usepackage{graphicx}
\newcommand{\calO}{{\cal O}}

\newcommand{\GeV}{{\rm GeV}}

\newcommand{\rmv}{{\rm v}}

\begin{document}
\baselineskip=16pt

\pagenumbering{arabic}

\vspace{1.0cm}

\begin{center}
{\Large\sf Unique neutrino mass operator at any mass dimension}
\\[10pt]
\vspace{.5 cm}

{Yi Liao\footnote{liaoy@nankai.edu.cn}}

{\it School of Physics, Nankai University, Tianjin 300071, China\\
Institut f\"ur Theoretische Physik, Universit\"at Heidelberg,
Germany\\
Center for High Energy Physics, Peking University, Beijing 100871,
China}

\vspace{2.0ex}

{\bf Abstract}

\end{center}

When the standard model is viewed as a low energy effective theory,
the neutrinos can obtain mass from higher dimensional operators. It
has been known for long that such an operator first appears at mass
dimension five and that it is unique. Here we show that the
effective neutrino mass operator at every higher dimension is
unique. This general claim is established using Young tableau, and
illustrated by exhausting all potentially different operators at
dimension seven. The result is relevant to the search of new physics
effects beyond neutrino mass that can arise at a relatively low
energy scale.

\begin{flushleft}


\end{flushleft}

\newpage

It has now been firmly established that neutrinos have
sub-electronvolt mass and that they mix in weak interactions. While
this provides the first piece of evidence for physics beyond
standard model (SM), the origin of such a small mass compared to
other particles that we know of has remained mysterious. It is
highly desirable that we should be able to observe some other
effects beyond the mass. This is however usually hard to achieve
since the new physics scale can easily exceed the accessibility of
experiments in the foreseeable future.

Indeed, when viewing SM as an effective field theory that
successfully incorporates phenomena at low energy below a hundred
GeV, there will be a tower of higher dimensional effective
interactions amongst known particles. These interactions are
suppressed compared to known interactions but will become more and
more important as we go up in the probe energy scale. The neutrino
mass can then be induced from such effective interactions. It has
been known for a long time \cite{Weinberg:1979sa} that the leading
interaction responsible for neutrino mass would arise from a unique
operator of mass dimension five
\begin{eqnarray}
\calO_5=\Big(\overline{F_L^C}\epsilon H\Big)\Big(H^T\epsilon
F_L\Big),
\end{eqnarray}
with an effective coupling constant $\lambda/\Lambda$, where
$\lambda$ is a product of coupling constants and $\Lambda$ a typical
mass scale in the underlying theory that induces the interaction.
Here $H$ is the SM Higgs field as a doublet in weak isospin,
$F_L=(\nu_L,\ell_L)^T$ is the left-handed lepton doublet field
composed of the neutrino $\nu$ and charged lepton $\ell$. The
superscript $C$ means charge conjugation, $\epsilon$ is the $2\times
2$ antisymmetric matrix in isospin space and the superscript $T$
denotes transpose in isospin space. Upon spontaneous symmetry
breaking when $H$ develops a vacuum expectation value (VEV), $\rmv$,
the operator $\calO_5$ yields a neutrino mass of order
$\lambda\rmv^2/\Lambda$. With the known $\rmv\sim 100~\GeV$ and
assuming the unknown $\lambda$ not much smaller than unity, the
sub-eV neutrino mass translates into a $\Lambda$ that is as high as
the grand unification scale.

It is the underlying theory producing the above effective
interaction that we are interested in. It was nicely observed some
years ago \cite{Ma:1998dn} that there are only three ways to realize
the interaction and that they correspond exactly to the three types
of seesaw mechanism suggested previously
\cite{type1,type2,Foot:1988aq}. To identify which of them is the
real physics for neutrino mass, we would have to observe other
effects. But this is hampered by the huge energy scale $\Lambda$
that is demanded by a tiny neutrino mass. To circumvent this
dilemma, one may attribute the occurrence of the dimension-five mass
operator $\calO_5$ to a higher order quantum effect that comes with
additional suppression factors \cite{Zee:1980ai}, or postpone the
appearance of neutrino mass operators to even higher dimensions that
will bring about more powers of $\rmv/\Lambda$ \cite{Babu:2009aq}.
The issue that we would like to address in this short note is, what
are the effective mass operators at a higher mass dimension when the
lowest dimension five operator is not available and is it
potentially possible to distinguish underlying theories by their
different contributions to various operators of same dimension? The
answer turns out to be surprisingly simple. At each mass dimension
$(2n+5)$ with $n$ being a nonnegative integer, there is a unique
mass operator of the form
\begin{eqnarray}
\calO_{(2n+5)}=\Big(\overline{F_L^C}\epsilon H\Big)\Big(H^T\epsilon
F_L\Big)\Big(H^\dagger H\Big)^n.
\end{eqnarray}
Although one can imagine to form out of isospin-half fields
arbitrarily high isospin states by tensor method in the intermediate
steps, they all condense into the basic invariant forms in terms of
two isospin-half fields. This simple feature can be attributed to
the fact that there are only two relevant fields both of which
belong to the fundamental representation of $SU(2)$. With this
uniqueness, it is still not possible to distinguish underlying
theories through their contributions to neutrino spectrum and mixing
even if the latter arises from a higher dimensional mass operator.
But with a large enough $n$ the chance to detect the new particles
would be much enhanced, making the underlying theory responsible for
$\calO_{(2n+5)}$ phenomenologically more viable.

Now we establish the above claim. The only fields in SM that are
relevant to neutrino mass are the above mentioned lepton field $F_L$
with isospin and hypercharge $(I,Y)=(1/2,-1)$ and the Higgs field
$H$ with $(I,Y)=(1/2,1)$, under the gauge group $SU(2)_L\times
U(1)_Y$. Since only the left-handed neutrino field is available, it
must be a Majorana particle when it is massive. We thus also need
the field $\overline{F_L^C}$ that transforms under gauge
transformations exactly as $F_L$. Our convention from now on is that
all fields are written as a column vector in isospin space though
$\overline{F_L^C}$ is a row-spinor in Dirac space. Also required is
the properly conjugated Higgs field $\tilde H=\epsilon H^*$ that
transforms like $H$ under isospin rotation but has an opposite
hypercharge. We consider how to form gauge invariant forms out of
these four fields that will yield a neutrino mass when $H$ develops
a VEV. This can be best solved by employing Young tableau for
$SU(2)$ in which gauge invariants appear as a two-row rectangle.

\begin{figure}
\begin{center}
\resizebox{0.9\textwidth}{!}{%
\includegraphics{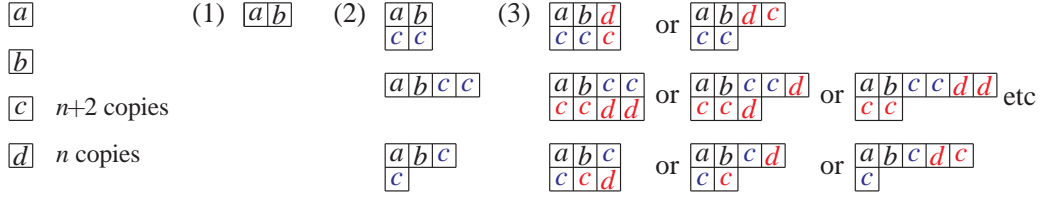}
} \caption{Young tableaux for operators. Notation:
$a=\overline{F_L^C}$, $b=F_L$, $c=H$, and $d=\tilde H$.}%
\label{fig1}
\end{center}
\end{figure}

Since the lepton number is violated by two units, the pair
$(F_L,\overline{F_L^C})$ must always appear once. To form an
$SU(2)$ invariant, we need an even number of the Higgs field; and
to balance the hypercharge we need two more $H$ than $\tilde H$.
Thus the dimension of mass operators jumps in the step of two.
With $n$ copies of $\tilde H$ one would form operators at
dimension $(5+2n)$. Although one can exhaust all independent
operators by multiplying the fields in any order (except that
$\overline{F_L^C}$ should appear on the left of $F_L$ to form a
Lorentz scalar), the easiest way, as will be clear later, is to
start with $(F_L,\overline{F_L^C})$ that have to appear anyway.
The potentially lowest-dimension operator with
$(F_L,\overline{F_L^C})$ alone does not actually exist since they
can only form an $SU(2)$ triplet, which is symmetric in isospin
space, and in addition, has a nonvanishing hypercharge. This is a
joint result of anticommutative fermionic fields and charge
conjugation, i.e., $\overline{\psi^C}\chi=\overline{\chi^C}\psi$.
With more than one generations of leptons one can form an isospin
singlet out of $(F_L,\overline{F_L^C})$ which however corresponds
to transitions between the neutral and charged leptons instead of
a neutrino mass. Our result thus holds true with any generations
of leptons, but we will not bother to attach a generation index.
This explains why in step (1) of fig. 1 the lepton fields are put
in the same row of the tableau. Putting them in the same column
would not give a contribution to mass.

Since there are two more $c=H$ than $d=\tilde H$ fields, we attach
the two $c$'s in step (2). When they are all put in the second row
as shown in the first tableau in step (2), we get the invariant at
dimension-5, which in our notation is, $\calO_5=2(ac)_0(cb)_0$,
where the subscript indicates the total isospin of the product in
the parentheses, e.g., $(cb)_0=(c^T\epsilon b)/\sqrt{2}$ in
isospin space. The remaining fields $H,\tilde H$ must appear in
pair in order to form invariants of a higher dimension. With one
pair of them, i.e., in step (3), one gets an invariant at
dimension-7, $\calO_7=2^{3/2}(ac)_0(cb)_0(dc)_0$, when they are
`correctly' put in the same column, or one has to wait until at
least the next pair to form an invariant when they are `wrongly'
put in the first row. With $n$ pairs of them, the lowest dimension
operator to appear is,
$\calO_{5+2n}=2^{1+n/2}(ac)_0(cb)_0[(dc)_0]^n$, corresponding to
the case when the pairs form a nonvanishing $2\times n$ rectangle
in whatever order. It is immaterial whether $c$ or $d$ in the same
column is put in the first row since each full column is an
invariant by itself; see later for an explicit demonstration of
the equivalence.

Now consider the second tableau formed in step (2). It is obvious
that at least two pairs of $(c,d)$ are required to form a two-row
rectangle. The first point to note is that the first box or the
first two boxes in the second row must be either left blank or
filled with $c$. If it is or at least one of them is filled with $d$
(these cases not depicted in fig .1), the tableau cannot develop
into an invariant at any later stage since in forming a rectangle
there will be at least one column that consists of two $c$'s thus
yielding zero. The cases when the first two boxes in the second row
are filled with $c$ are shown in fig. 1. The first one gives the
dim-9 operator, $\calO_9$, while the other two (plus the cases not
shown with a one-$c$ box or blank second row) will develop into an
invariant in later steps when no column contains only $c$ or $d$.
All these higher dimensional operators have the unique form of
$\calO_{5+2n}$. From this point on, there will be nothing new with
the third tableau in step (2) since all possibilities have been
exhausted, and the procedure obviously generalizes to any odd
dimensions.

We show by an explicit example of Young tableau that reordering the
multiplication of fields when forming an invariant will not change
the invariant, up to a sign arising from interchanging fermionic
fields and phase conventions in composing eigenstates of isospin. It
suffices to examine the lowest dimension operator $\calO_5$ since
each additional column will multiply as a separate invariant
$(dc)_0$. Denoting as usual the Young tableau by its content from
upper to lower rows and from left to right in each row, the content
of a tableau can be translated into an eigenfunction formed with the
fundamental isospin-half spinors. For example,
\begin{eqnarray}
\{\{a,b\},\{c,c\}\}=(ac)_0(bc)_0
=\frac{1}{2}(a_+c_--a_-c_+)(b_+c_--b_-c_+),
\end{eqnarray}
where the subscript $+$ ($-$) denotes the upper (lower) component of
a spinor, e.g., $\nu_L$ ($\ell_L$) in $F_L$, while
\begin{eqnarray}
\{\{a,c\},\{c,b\}\}=(ac)_0(cb)_0
=\frac{1}{2}(a_+c_--a_-c_+)(c_+b_--c_-b_+)=-\{\{a,b\},\{c,c\}\}.
\end{eqnarray}
This also offers an example showing that the two boxes in the same
column are interchangeable.

Finally, we illustrate our general result by constructing explicitly
all potentially independent operators at dimension-7 in terms of
Clebsch-Gordan coefficients. We continue to use the field notations
in fig. 1. We want to form isospin invariants with one copy of
$a,~b,~d$ and three copies of $c$. Since $(ab)$ and $(cc)$ exist
only with isospin-1, there are following possibilities to start with
that may develop into an invariant:
\begin{eqnarray}
\textrm{(i)}:~(ab)_1(cd)(cc)_1;~\textrm{(ii)}:~(ad)(bc)(cc)_1;
~\textrm{(iii)}:~(ac)(bd)(cc)_1; ~\textrm{(iv)}:~(ac)(bc)(cd).%
\label{eq_list}
\end{eqnarray}
Here the product without an isospin index means that it may exist in
either isospin 0 or 1 states. To proceed further, we notice that a
product of two isospin-1 states must be in a state of isospin not
exceeding one if it is about to develop into an invariant with a
third state of isospin not larger than one. We continue to denote
multiple isospin composition by parentheses with a subscript. For
case (i), there are two possibilities corresponding to two possible
isospins of $(cd)$. The composition is simple when $(cd)$ has
isospin 0:
\begin{eqnarray}
({\bf 1})_{\textrm{(i)},1}&=&
\left((ab)_1(cc)_1\right)_0(cd)_0\nonumber\\
&=&\frac{1}{\sqrt{6}}\big[a_+b_+c_-c_- -(a_+b_-+a_-b_+)c_+c_-
+a_-b_-c_+c_+\big](c_+d_--c_-d_+),
\end{eqnarray}
which is $\calO_7/\sqrt{6}$ in our standard operators. Although the
intermediate steps are complicated for $(cd)_1$, the result is
simple
\begin{eqnarray}
({\bf 1})_{\textrm{(i)},2}&=&\left((ab)_1(cd)_1(cc)_1\right)_0
=-\frac{1}{\sqrt{2}}({\bf 1})_{\textrm{(i)},1}.
\end{eqnarray}
The difference in normalization just reflects the fact we started
with an arbitrary normalization in the list (\ref{eq_list}) of
cases. For case (ii) there are only three possibilities since $(ad)$
and $(bc)$ cannot stay simultaneously in isospin-0 states to form an
invariant with $(cc)_1$ later on. Further, it is not possible to
construct an invariant $\left((bc)_1(cc)_1\right)_0$ due to multiple
occurrence of identical $c$'s. Skipping the details, the result is
\begin{eqnarray}
({\bf 1})_{\textrm{(ii)},1}&=&\left((ad)_1(cc)_1\right)_0(bc)_0
=({\bf 1})_{\textrm{(i)},1},\nonumber\\
({\bf 1})_{\textrm{(ii)},2}&=&\left((ad)_1(bc)_1(cc)_1\right)_0
=-\frac{1}{\sqrt{2}}({\bf 1})_{\textrm{(i)},1}.
\end{eqnarray}
The case (iii) is related to (ii) by interchanging $c$ and $d$ in
factors other than $(cc)_1$. Case (iv) has apparently most
possibilities with the nonvanishing ones being
\begin{eqnarray}
({\bf 1})_{\textrm{(iv)},1}&=&
(ac)_0\big((bc)_1(cd)_1\big)_0=-\frac{1}{2}({\bf 1})_{\textrm{(i)},1},
\nonumber\\
({\bf 1})_{\textrm{(iv)},2}&=&
\big((ac)_1(cd)_1\big)_0(bc)_0=-\frac{1}{2}({\bf 1})_{\textrm{(i)},1},
\nonumber\\
({\bf 1})_{\textrm{(iv)},3}&=&
\big((ac)_1(bc)_1\big)_0(cd)_0=-\frac{1}{2}({\bf 1})_{\textrm{(i)},1},
\nonumber\\
({\bf 1})_{\textrm{(iv)},4}&=&
(ac)_0(bc)_0(cd)_0=\frac{\sqrt{3}}{2}({\bf 1})_{\textrm{(i)},1}.
\end{eqnarray}

The tiny neutrino mass signals the existence of new physics beyond
standard model. At low energy it can be described as a consequence
of high dimensional effective interactions amongst standard model
particles. It was known that such an interaction may first appear at
dimension five and that it is unique. We have shown in this work
that all those interactions of dimension higher than five are also
unique at each dimension. These interactions are called for in new
physics models to relax the tension between the tiny neutrino mass
and an accessible new physics scale by forbidding the lowest
dimension five operator. The uniqueness significantly simplifies the
analysis of neutrino mass and mixing in those models. We emphasize
that this uniqueness is restricted to the minimal case of one Higgs
doublet; for instance, with two Higgs doublets the number of mass
operators increases quickly with their dimension \cite{Liao}.

\vspace{0.5cm}
\noindent %
{\bf Acknowledgement}

This work is supported in part by the grants NSFC-10775074,
NSFC-10975078, and the 973 Program 2010CB833000. I would like to
thank T. Plehn for his kind invitation for a visit and Institut
f\"ur Theoretische Physik, Universit\"at Heidelberg for hospitality.
I thank Tianjun Li and Zhi-Zhong Xing for helpful discussions.

\noindent %


\begin{thebibliography}{100}

\bibitem{Weinberg:1979sa}
  S.~Weinberg,
  Phys.\ Rev.\ Lett.\  {\bf 43}, 1566 (1979).

\bibitem{Ma:1998dn}
  E.~Ma,
  Phys.\ Rev.\ Lett.\  {\bf 81}, 1171 (1998)
  [arXiv:hep-ph/9805219].

\bibitem{type1}M. Gell-Mann, P. Ramond, R. Slansky, in:
D. Freedman, P. van Nieuwenhuizen (Eds.), Supergravity,
North-Holland, Amsterdam, 1979, p.315; T. Yanagida, in: O. Sawada,
A. Sugamoto (Eds.), Proceedings of the Workshop on Unified Theory
and Baryon Number in the Universe, KEK, Japan, 1979; R.N. Mohapatra,
G. Senjanovic, Phys. Rev. Lett. {\bf 44} (1980) 912.

\bibitem{type2}
  W.~Konetschny and W.~Kummer,
  Phys.\ Lett.\  B {\bf 70}, 433 (1977);
  T.~P.~Cheng and L.~F.~Li,
  Phys.\ Rev.\  D {\bf 22}, 2860 (1980);
  J.~Schechter and J.~W.~F.~Valle,
  Phys.\ Rev.\  D {\bf 22}, 2227 (1980).

\bibitem{Foot:1988aq}
  R.~Foot, H.~Lew, X.~G.~He and G.~C.~Joshi,
  Z.\ Phys.\  C {\bf 44}, 441 (1989).

\bibitem{Zee:1980ai}
  A.~Zee,
  Phys.\ Lett.\  B {\bf 93}, 389 (1980)
  [Erratum-ibid.\  B {\bf 95}, 461 (1980)];
  A.~Zee,
  Nucl.\ Phys.\  B {\bf 264}, 99 (1986);
  K.~S.~Babu,
  Phys.\ Lett.\  B {\bf 203}, 132 (1988).

\bibitem{Babu:2009aq}
For recent discusions on models that induce neutrino mass by higher
dimensional operators, see, e.g.:
  K.~S.~Babu, S.~Nandi and Z.~Tavartkiladze,
  Phys.\ Rev.\  D {\bf 80}, 071702 (2009)
  [arXiv:0905.2710 [hep-ph]];
  Z.~z.~Xing and S.~Zhou,
  Phys.\ Lett.\  B {\bf 679}, 249 (2009)
  [arXiv:0906.1757 [hep-ph]];
  F.~Bonnet, D.~Hernandez, T.~Ota and W.~Winter,
  JHEP {\bf 0910}, 076 (2009)
  [arXiv:0907.3143 [hep-ph]];
  I.~Picek and B.~Radovcic,
  Phys.\ Lett.\  B {\bf 687}, 338 (2010)
  [arXiv:0911.1374 [hep-ph]].

\bibitem{Liao}
Y.~Liao, Neutrino mass operators of dimension up to nine in
two-Higgs-doublet model, to appear.

\end{thebibliography}
\end{document}